# Transition to a weaker Sun: Changes in the solar atmosphere during the decay of the Modern Maximum

K. Mursula[1,2], A. A. Pevtsov[2], T. Asikainen[1], I. Tähtinen[1], and A. R. Yeates[3]

[1] Space Climate Group, Space Physics and Astronomy Res. Unit, University of Oulu, 90014 Oulu, Finland
  e-mail: `kalevi.mursula@oulu.fi`
[2] National Solar Observatory, 3665 Discovery Drive, 3rd Floor, Boulder, CO 80303, USA
[3] Department of Mathematical Sciences, Durham University, Durham, UK

March 12, 2024

**ABSTRACT**

The Sun experienced a period of unprecedented activity during the 20th century, now called the Modern Maximum (MM). The decay of the MM after its maximum in cycle 19 has changed the Sun, the heliosphere, and the planetary environments in many ways. However, studies disagree on whether this decay has proceeded synchronously in different solar parameters or not. One of the related key issues is if the relation between two long parameters of solar activity, the sunspot number and the solar 10.7 cm radio flux, has remained the same during this decay. A recent study argues that there is an inhomogeneity in the 10.7 cm radio flux in 1980, which leads to a step-like jump ("1980 jump") in this relation. If true, this result would reduce the versatility of possible long-term studies of the Sun during the MM. Here we aim to show that the relation between sunspot number and 10.7 cm radio flux does indeed vary in time, not due to an inhomogeneous radio flux but due to physical changes in the solar atmosphere. We used radio flux measurements made in Japan at four different wavelengths, and studied their long-term relation with the sunspot number and the 10.7 cm radio flux during the decay of MM. We also used two other solar parameters, the MgII index and the number of solar active regions, in order to study the nature of the observed long-term changes in more detail. We find that the 1980 jump is only the first of a series of 1-2-year "humps" that mainly occur during solar maxima. All five radio fluxes depict an increasing trend with respect to the sunspot number from the 1970s to 2010s. These results exclude the interpretation of the 1980 jump as an inhomogeneity in the 10.7 cm flux, and reestablish the 10.7 cm flux as a homogeneous measure of solar activity. The fluxes of the longer radio waves are found to increase with respect to the shorter waves, which suggests a long-term change in the solar radio spectrum. We also find that the MgII index of solar UV irradiance and the number of active regions also increased with respect to the sunspot number, further verifying the difference in the long-term evolution in chromospheric and photospheric parameters. Our results provide evidence for important structural changes in solar magnetic fields and the solar atmosphere during the decay of the MM, which have not been reliably documented so far. We also emphasize that the changing relation between the different (e.g., photospheric and chromospheric) solar parameters should be taken into account when using the sunspot number or any single parameter in long-term studies of solar activity.

**Key words.** Solar activity – Solar radio emissions – Space climate

## 1. Introduction

The Sun has experienced large variations in its magnetic activity during the last 100 years. Solar activity reached unprecedentedly high levels in the mid-20th century (Usoskin et al. 2003; Solanki et al. 2004; Hathaway 2015), during a period now called the Modern Maximum (MM), which was the most recent cycle of the roughly centennial Gleissberg cyclicity (Gleissberg 1939; Ogurtsov et al. 2002). After its maximum in solar cycle 19 (SC19), solar activity considerably decreased until cycle 24, which attained roughly the same low level as cycles one century ago, before the MM onset. Weak solar activity culminated in an exceptionally long and deep minimum between cycles 23 and 24. The ongoing cycle 25 is now known to be somewhat more active than cycle 24 (for predictions, see, e.g., Pesnell (2020); Petrovay (2020); Bhowmik et al. (2023); Upton & Hathaway (2023)); for the current status of cycle 25, readers are referred to SILSO web page[1]). This roughly 50-year time interval from SC19 to SC24 of weakening solar activity forms the decay phase of the MM. It is interesting to note that the start in 1957 of the space age (the period of satellite missions) coincides with the peak of the MM. Accordingly, all satellite observations about the near-Earth space are from the maximum and decay period of the MM, which may have given us a somewhat one-sided view of long-term solar activity.

The decrease in overall solar activity during the decay of the MM has changed the Sun, the heliosphere, and the planetary environments in a versatile and complex way. The basic parameter used to measure the magnetic activity of the Sun, the sunspot number, reduced from its all-time maximum of 285.0 (according to international sunspot number, version 2) in SC19 to 116.4 at the maximum of SC24. Accordingly, the amplitudes of sunspot cycles have reduced by a factor of about 2.4 during the decay of the MM.

Sunspots are an important (although not the only) source of the global photospheric magnetic field. Thus the decreasing number of sunspots has also affected the magnetic field on the whole solar surface. The new magnetic flux appearing on the solar surface as sunspots or other active regions forms surges that transport magnetic flux to high solar latitudes, affecting the strength of solar polar fields. With the decreasing number of sunspots (and other magnetic elements), the strength of polar fields has reduced (Wang et al. 2009) during the last four cycles.

---
[1] https://www.sidc.be/SILSO/

Weaker global and polar magnetic fields have also reduced the intensity of the heliospheric magnetic field (HMF) (Smith & Balogh 2008; Virtanen & Mursula 2019), although the reduction in the HMF, that is, in the open solar magnetic field, is smaller than expected based on the weakening photospheric fields (Wang et al. 2000; Virtanen et al. 2020). This mismatch is an urgent research topic that is still unresolved (Petrie 2015; Linker et al. 2017). Weaker polar fields have also widened the heliospheric current sheet and the streamer belt (Mursula et al. 2022), reducing the occurrence of high-speed solar wind streams at the Earth. Moreover, the widening of the current sheet has also increased the number of low-latitude coronal holes that were particularly persistent during the declining phase of SC23 (Gibson et al. 2009; Fujiki et al. 2016; Hamada et al. 2021).

The reducing number of sunspots has also produced a smaller number of coronal mass ejections. As a consequence the number of intense geomagnetic storms has reduced from SC19 to SC24, roughly by the same factor as the sunspot cycle heights have reduced (Mursula et al. 2022). Due to the wider heliospheric current sheet, the Earth now spends more than half of the time within the slow solar wind. As a consequence, the geomagnetic disturbance level has considerably reduced during the last 20 years. The year 2009 was the quietest year during the whole space age both in overall geomagnetic activity, as measured, for example, by the Kp/Ap, aa, and other similar indices (Echer et al. 2012; Mursula et al. 2017; Sumaruk et al. 2023), and in the number of magnetic storms (Mursula et al. 2022; Owens et al. 2022).

The intensity of total solar electromagnetic radiation (total solar irradiance, TSI) follows the sunspot number quite closely over the measured time interval (Fröhlich 2013). Accordingly, the amplitude of the TSI cycle, and the average TSI have reduced since cycle 23 (Kopp 2016; Chatzistergos et al. 2023). However, there is no firm evidence of a possible small long-term change in the minimum-time TSI level (Dudok de Wit et al. 2017; Tebabal et al. 2017; Yeo et al. 2017; Lean 2018; Chatzistergos et al. 2020; Montillet et al. 2022). Similarly, the brightnesses (irradiances) of most spectral emission bands, except for near-infrared, roughly follow the evolution of sunspots over the solar cycle (Pagaran et al. 2011; Ermolli et al. 2013; Yeo et al. 2015).

Accordingly, all solar, heliospheric, and even solar-terrestrial parameters seem to respond to the decreasing solar activity during the decay of the MM in a seemingly similar way. However, these changes may still not proceed synchronously in all parameters. Changes may well be somewhat different, for example, between parameters that differ in how they depend on solar magnetic fields, or that have sources at different layers of the solar atmosphere. Thus, a study of possible long-term changes in the relation between different solar parameters or in the structure of solar magnetic fields is well motivated, although quite challenging.

One possibility is to study the detailed long-term consistency of several solar parameters. For such a consistency study, one needs at least two reliable, accurate and homogeneous long-term parameters. Sunspot number is an obvious candidate for one, and the revised (version 2) sunspot number (Clette & Lefèvre 2016) is supposed to be homogeneous over a longer time interval than the original version. The solar 10.7 cm radio flux is another long-term (about 75 years) solar parameter, which has generally been considered to be homogeneous over its whole measurement interval (see, e.g., Tapping 2013).

Accordingly, one basic question related to the long-term consistency is if the mutual relation between the two longest and most homogeneous parameters of solar activity, the sunspot number and the solar 10.7 cm radio flux, has changed or remained the same during the decay of the MM. A number of studies have found that this relation has indeed changed during the last decades (Lukianova & Mursula 2011; Tapping & Valdés 2011; Tapping & Morgan 2017; Bruevich & Bruevich 2019; Laštovička & Burešová 2023).

While most studies agree on a changing relation between these two parameters, no consensus on the actual validity of this change (whether real or artificial), its cause or even on its timing has been reached among the different studies. Recently a detailed study was published (Clette 2021) which argued that the change in the mutual relation between sunspot number and the 10.7 cm radio flux is due to a change in the post-processing procedure of the radio flux, which produced a step-like jump in this relation from 1980 to 1981 (so-called 1980 jump). If this result was valid, the 10.7 cm radio flux would be temporally inhomogeneous, which would prohibit its use as one of the longest parameters in studies of solar long-term relations. This would reduce the versatility of possible studies of the long-term evolution of the Sun during the decay of the MM.

Here we use the 10.7 cm radio flux measured in Canada and four Japanese radio flux measurements that are completely independent of the 10.7 cm radio flux in order to study their relation with the sunspot number during the peak and decay of MM since 1952. We can exclude the claim that the 1980 jump is due to an inhomogeneity in the 10.7 cm flux. We find that the 1980 jump is not a unique step-like increase, but the first in a series of short-term (1-2 years) humps which make all five radio fluxes increase with respect to the sunspot number from the late 1970s until 2010s. We also find a long-term change between the longer and shorter radio wavelengths, which suggests an interesting long-term change in the solar spectrum at radio frequencies. In addition to solar radio fluxes, also the MgII index of solar UV irradiance and the number of active regions increased with respect to the sunspot number at the same time. We argue that our results give evidence for important structural changes in solar magnetic fields and solar atmosphere during the decay of the MM.

This paper is organized as follows. We introduce the Canadian (Penticton) 10.7 cm radio flux and the four Japanese (Toyokawa/Nobeyama) radio fluxes in Section 2, and sunspot number, solar MgII index and the database of solar active regions in Section 3. In Section 4 we analyze the claimed 1980 jump in the mutual relation between sunspot number and the 10.7 cm radio flux. In Section 5 we study the relation between sunspot number and five different radio fluxes. The relation between the Canadian 10.7 cm radio flux and the four Japanese radio fluxes is examined in Section 6. In Section 7 we study the mutual relation between sunspot number and two different solar parameters, the MgII index and the number of solar active regions. In Section 8 we discuss the results obtained in this work, and note on connections to a few other related studies. We also show that the small nonlinearity in the relation between the yearly values of different parameters has practically no effect to our results. We also give our interpretation on the observed results. Finally in Section 9 we give our conclusions from this study.

## 2. Data: Radio emissions

### 2.1. Penticton 10.7 cm radio flux

Solar radio emission bursts were detected soon after the World War II in the United Kingdom (Hey 1946) and, subsequently, solar radio monitoring observations started at a few observatories in the late 1940s (see, e.g., Tanaka et al. 1973; Tapping 2013). Continuous measurements of solar 10.7 cm wavelength (2.8 GHz frequency) radio emissions have been made in Canada since February 14, 1947 (Covington 1947, 1948; Tapping 2013). The location of these measurements has moved twice, from Ottawa to Algonquin in 1962 and then to Dominion (Penticton, British Columbia) in 1991 (Tapping 2013). On most days, three measurements were made, out of which the noon measurement was preferably used.

Solar radio fluxes have two versions that are typically distributed at all data servers: the Observed flux and the Adjusted flux. The Observed flux is the original, measured flux, which includes the influence of the annually varying distance of the Earth from the Sun. Therefore, the Observed flux is normally used, for example, when studying solar effects to the Earth. The Adjusted flux is the flux normalized to the constant radial distance of one astronomical unit (1AU) and is therefore the correct index when studying the Sun itself, as in this study.

The daily 10.7 cm radio flux data was obtained from the LISIRD database[2], which provides a link to the NOAA version of the Penticton (Canadian) F10.7 index[3]. The NOAA F10.7 index covers the time from the start of continuous 10.7 cm measurements until the end of April, 2018, when the NOAA stopped the index production. However, we will use this index only from 1952 onward in order to have a common time interval for all radio observations. We continued the NOAA F10.7 index from May 2018 onward using the recent Penticton radio flux data available from the NRCan server, as described in more detail in Mursula (2023). This produces a homogeneous F10.7 index from 1952 to 2022, which is used in this study.

The daily and yearly mean fluxes of 10.7 cm radio emissions are depicted in the third panel of Fig. 1. One can see the large scatter of daily F10.7 values around the yearly means. Figure 1 also depicts the large solar cycle variation of centimetric radio emissions, in good agreement with sunspots (depicted in the bottom panel of Fig. 1). Their correlation is known already from the early years of radio flux measurements (Covington 1947; Lehany & Yabsley 1948). In addition to the sunspot cycle variation, one can see that the differences in the heights of the six cycles covered by 10.7 cm observations are also very similar to those of sunspot cycles.

Thus, the time series included in Fig. 1 depict the common view that the solar 10.7 cm radio flux has followed the sunspot number during the last 6-7 cycles very closely, despite the large variation of solar activity during this time. Accordingly, those changes in the mutual relation between the 10.7 cm radio flux and the sunspot number that have been found in a few studies (see, e.g., Lukianova & Mursula 2011; Tapping & Morgan 2017; Bruevich & Bruevich 2019; Clette 2021) cannot be seen directly in the time series but require a more detailed examination.

### 2.2. Nobeyama radio fluxes

Solar radio flux observations in Japan started in the early 1950s (for a recent review, see Shimojo & Iwai (2023)). Measurements were first, since 1951, made in Toyakawa with 3.75 GHz (8 cm) waves, later in the 1950s they started measurements in three other frequencies as well. Continuous observations were made at these four frequencies (1, 2, 3.75, and 9.4 GHz; corresponding wavelengths 30 cm, 15 cm, 8 cm, and 3.2 cm) in Toyakawa until 1994. Since then, observations at these frequencies are continued in Nobeyama by the National Astronomical Observatory of Japan, together with three additional bands of higher frequency (17, 35, and 80 GHz; corresponding wavelengths 17.6 cm, 8.6 mm, and 3.7 mm), which were installed in the late 1970s and early 1980s (Nakajima et al. 1985).

We used here the four longest-measured wavelengths of Japanese radio flux observations. Panels 1, 2, 4, and 5 of Figure 1 (ordered according to wavelength and called F30, F15, F8, and F3.2) depict the daily and yearly mean fluxes of these four radio emissions. We note that the data intervals of the four Japanese radio fluxes vary slightly. We used the F8 flux in its full years from 1952 to 2021. The three other lines start in 1957 and extend until 2021, except for F15 which has a long data gap around 2018, and is therefore limited to extend only until 2017.

These four radio fluxes measured in Japan, together with the Penticton 10.7 cm radio flux form a unique database of solar observations that are useful, for example, for long-term solar cycle studies (Shimojo et al. 2017; Tapping & Morgan 2017). Figure 1 shows that all these five radio fluxes vary roughly in phase with the sunspot number over the solar cycle and even depict quite similar height differences between the six cycles included in the study. We note that the radio fluxes attain nonzero minima at solar minima due to thermal emission. These minima, naturally, decrease with increasing wavelength according to the solar spectrum. However, the solar cycle variation is relatively largest (about ±50% around the mean in yearly values) for the three longest wavelengths (F30, F15, F10.7), slightly smaller for F8 and considerably smaller (about ±20%) for the shortest wavelength F3.2.

## 3. Data: Other solar parameters

### 3.1. Sunspot number

We use the daily, monthly and yearly total sunspot number of version 2 (Clette et al. 2015, 2016)) served at the World Data Center SILSO, Royal Observatory of Belgium[4]. We have depicted the daily and yearly mean sunspot number in 1952-2021 in the bottom panel of Fig. 1, and in 1977-2021 in the bottom panel of Fig. 2 for the comparison with the other parameters.

### 3.2. Solar MgII core to wing ratio

The core to wing ratio of the Magnesium-II doublet at about 280 nm (MgII index) is used as a standard proxy measure for solar UV-EUV irradiance (see, e.g., Deland & Cebula 1993; Viereck & Puga 1999; Viereck et al. 2001, 2004; Snow et al. 2014). MgII index is an index of overall chromospheric activity and has a connection to solar plages. Magnesium-II measurements from several satellites (GOME, SCIAMACHY, GOME-2A, and GOME-2B) have been used to construct a long-term Mg II index, the so-called Bremen MgII composite index[5] updated by M. Weber. The daily and yearly means of the Bremen MgII composite index in 1979-2021 are depicted in the top panel of Fig. 2. One can see that the MgII index follows the sunspot cycle quite well, but the differences between the two are slightly more conspicuous than in the case of the radio fluxes and the sunspot number. We also note that the relative solar cycle variation of the MgII index is much smaller than at radio frequencies, only about ±6% around the mean in yearly values. However, this is larger than the solar cycle amplitude of other UV frequencies above 240 nm (not shown here).

---

[2] https://lasp.colorado.edu/lisird/
[3] https://lasp.colorado.edu/lisird/data/noaa_radio_flux
[4] https://www.sidc.be/silso/
[5] https://www.iup.uni-bremen.de/gome/gomemgii.html

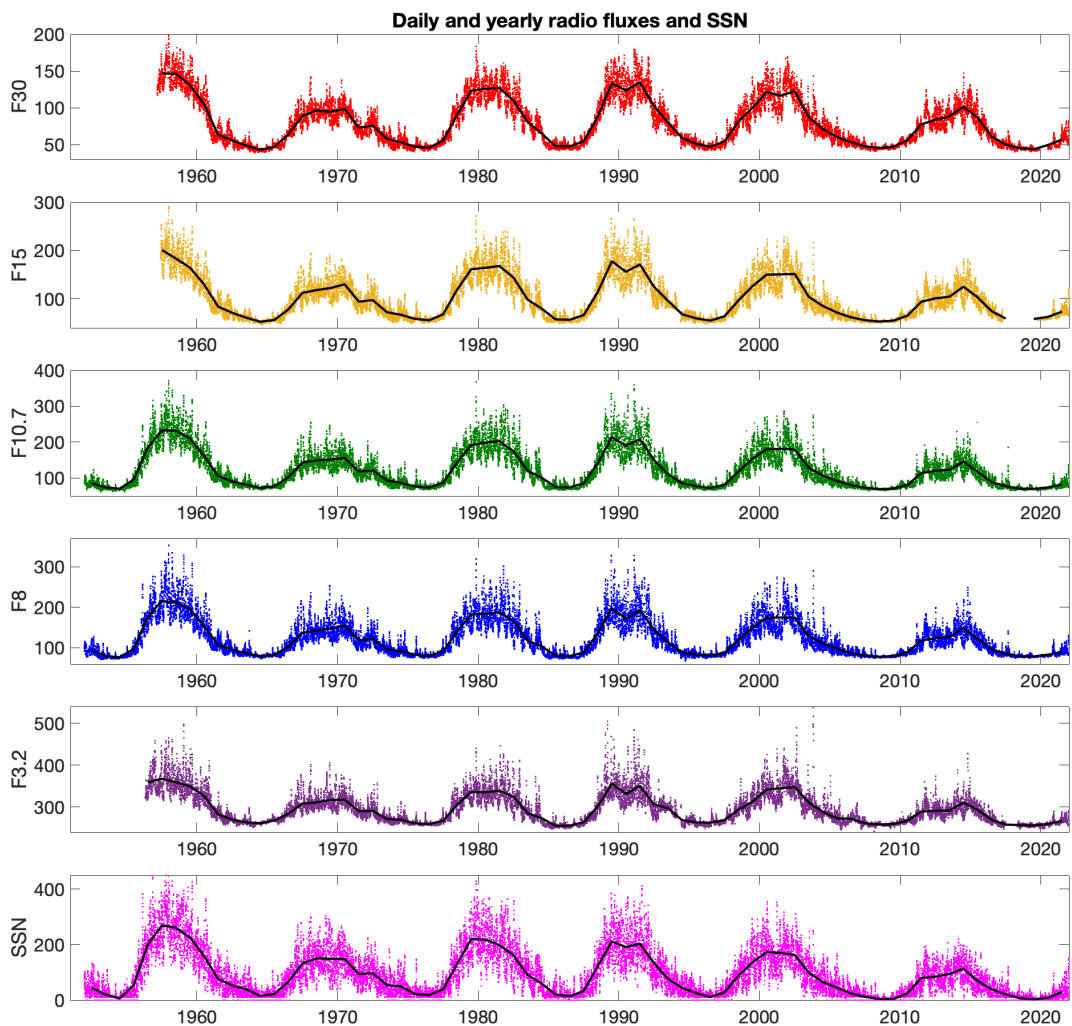

**Fig. 1.** Daily (colored dots) and yearly (black lines) means of radio fluxes at five different wavelength bands from top to bottom: F30 cm (red), F15 cm (yellow), F10.7 cm (green), F8 cm (blue), and F3.2 cm (purple). Daily (magenta dots) and yearly (black line) sunspot number are depicted in the bottom panel.

### 3.3. Active regions

While sunspots are quite well defined, an "active region" is a less clearly specified concept, whose meaning depends on the context and requires a detailed definition of related data and methods. The formation of a sunspot requires a very strong field of about 1500 G. However, observations measuring the photospheric magnetic field over (nearly) the whole solar disk typically find many more magnetically active regions with a weaker magnetic field intensity than in sunspots.

Active regions can be defined by imposing a minimum field intensity threshold, which would depend, for example, on the sensitivity and resolution of the instrument. Typical threshold values for active regions determined from photospheric magnetic field observations are on the order of several tens of Gauss, that is, much less than the field intensity needed for a sunspot to form. So-defined active regions include, for example, photospheric faculae, which have close counterparts in chromospheric plages that, on the other hand, can be defined as active regions based on various emission lines, such as the CaII K line.

The intensity of the magnetic field and the density of solar plasma play significant roles in physical mechanisms that produce electromagnetic emissions at the measured radio frequencies. Solar 10.7 cm radio emissions contributing to the varying, activity-dependent S-component are mainly produced in the solar chromosphere and lower corona by three types of physical mechanisms: electron bremsstrahlung (also called thermal free-free) emission, electron thermal gyroresonance, and nonthermal emissions (Krueger 1979; Tapping 1987; Tapping & Detracey 1990). Since the magnetic fields in sunspots are much stronger than in non-sunspot active regions, the electrons in sunspot regions emit most of the gyroresonance radiation contributing to the 10.7 cm wavelength band. Bremsstrahlung is the dominant emission in the less intense active regions, in particular in plages. Moreover, since plages typically cover a much larger area on the solar disk than simultaneous sunspots, by far most of the 10.7 cm radio emissions are produced in the non-sunspot active regions by bremsstrahlung. On average, only less than 10% of the variable 10.7 cm flux has been estimated to come from sunspots, while more than 60% comes from non-sunspot active regions (Schonfeld et al. 2015; Tapping & Detracey 1990). However, since the occurrence of plages follows the sunspot occurrence (sunspot cycle) relatively closely (Bertello et al. 2016; Chatzistergos et al. 2022; Nèmec et al. 2022a,b), radio flux and sunspots also follow each other relatively closely, at least on monthly and longer timescales (Covington 1947; Lehany & Yabsley 1948; Johnson 2011; Tiwari & Kumar 2018; Clette 2021).

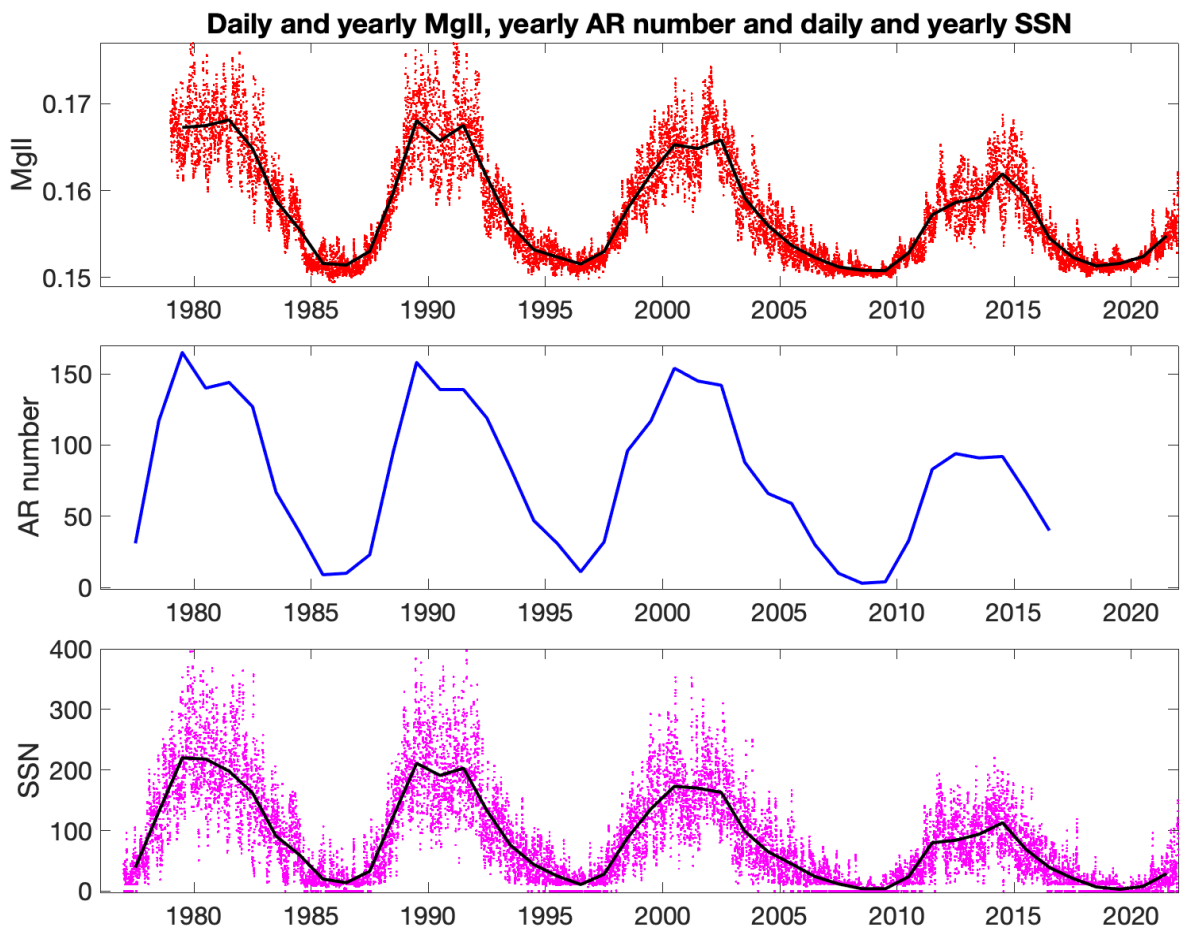

**Fig. 2.** Time series of three solar parameters used in this study. Top: Daily (red dots) and yearly (black line) means of Bremen MgII composite index for 1979-2021. Middle: yearly number of active regions for 1977-2016 (blue line). Bottom: Daily (magenta dots) and yearly (black line) sunspot number for 1977-2021.

Here we will use active regions derived from the synoptic Carrington maps of the photospheric magnetic field line-of-sight observations measured by the National Solar Observatory (NSO) Kitt Peak Vacuum Telescope (KPVT) (Livingston et al. 1976) and Synoptic Optical Long-term Investigations of the Sun (SOLIS) (Keller et al. 2003; Gosain et al. 2013; Balasubramaniam & Pevtsov 2011) instruments. The derivation of these active regions is described in Whitbread et al. (2018). Active regions were identified from the Carrington maps using an automated procedure by a code available at github[6]. Each active region is defined as a region of connected pixels with magnetic flux density above a threshold of 39.8 G. Regions were not assumed to be bipolar but can have a complex, multipolar structure and field distribution. There was no tracking of active regions, and persistent active regions may appear in several Carrington maps.

Active region data were taken from Yeates (2016). The active region list starts in Dec 6, 1976 (Carrington rotation 1649) and extends until Oct 12, 2017 (CR 2196). We use here only the yearly mean number of active regions for fully covered years from 1977 to 2016. They are depicted in the second panel of Fig. 2.

## 4. Analysis of the 1980 jump

As mentioned above, Clette (2021) suggested that there is a step-like change from 1980 to 1981 in the long-term relation between the F10.7 flux and the sunspot number, which is depicted in their Fig. 24. We have depicted a similar (but not identical) plot in the left panel of Fig. 3. For that, monthly F10.7 fluxes were first correlated with monthly sunspot numbers, and sunspot numbers scaled to the F10.7 flux level (to be called here the correlated sunspot numbers) were obtained from the best-fit line of this correlation. Then monthly ratios between the F10.7 fluxes and the correlated sunspot numbers were calculated and plotted. The left panel of Fig. 3 shows these ratios (as red dots in 1979-1980 and blue dots in 1981-1982) during the four years from 1979 to 1982, in a similar was as presented by Clette (2021). The right panel of Fig. 3 shows the same ratios (small blue dots) for the whole time interval studied here from 1952 to 2021.

---
[6] https://github.com/antyeates1983/sft_data

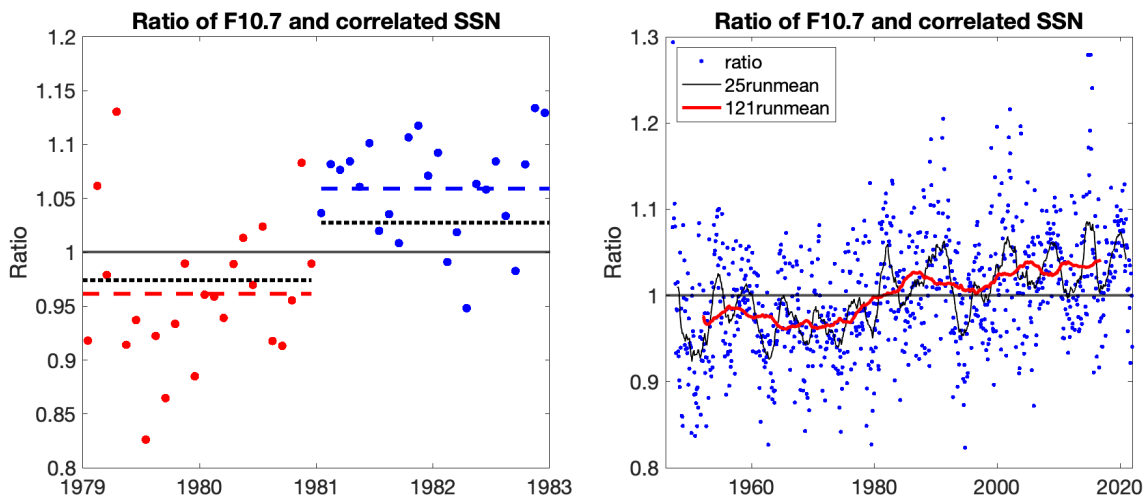

**Fig. 3.** Ratios between the monthly F10.7 fluxes and the correlated sunspot numbers. Left: Ratios for 1979-1982, with ratios in 1979-1980 depicted as red dots and in 1981-1982 as blue dots (cf. Fig. 24 of Clette (2021)). Red and blue dashed lines denote the means of ratios (dots) over the corresponding 2-year periods. Two black dotted lines depict the overall means of ratios over 1947-1980 and 1981-2015. Right: Same monthly ratios (blue dots) for the time interval 1947-2021. Black and red curves denote the 25-month (roughly two-year) and 121-month (roughly 10-year) running means of the monthly ratios.

We have also plotted in the left panel of Fig. 3 the means of these monthly ratios for the two 2-year intervals (1979-1980 and 1981-1982) as dashed lines in the same color as the dots in the respective intervals, as well as the means of the whole pre-1981 time interval and the whole post-1981 time interval as two black dotted lines. As seen in the left panel of Fig. 3, the blue line (the mean over 1981-1982) is far above the black line (the whole post-1981 interval mean). Also, the red line (the mean over 1979-1980) is clearly further away from unity than the pre-1981 mean. The difference between the two 2-year means is almost twice larger than the difference between the means for the post-1981 and pre-1981 intervals.

We note that the colored lines in Fig. 24 of Clette (2021) are not constructed as the means of ratios over the respective 2-year time intervals. (The description of Fig. 24 in Clette (2021) is rather unclear). Rather, they are constructed from the slopes of three correlation lines, one being the global fit to the data, with SSN limited to within 25-290 (slope given in their Table 4), the two other being separate fits to the data before and after 1980 (slopes in their Table 9). Clette (2021) calculated the ratio of the post-1980 slope to the global slope (which is larger than one) and, similarly, the ratio of the pre-1980 slope to the global slope (smaller than one), and used those when plotting the two colored lines over the two 2-year intervals. However, Clette (2021) slightly lowered these ratios in Fig. 24 so that their mutual difference (estimate of the jump) remained the same but they were placed symmetrically around unity (their mean becoming one). In this way, Clette (2021) gave the impression of a very large (about 10%) permanent step-like change across 1981, calling this the 1980 jump. This step was, without any additional evidence, suggested to be due to a learning period when a new post-processing method was implemented by the new team after the retirement in 1979 of A. E. Covington, who had directed the F10.7 flux production continuously since 1947.

The right panel of Fig. 3 (note the slightly larger range in this panel) sets the increase across 1980/1981 into a longer perspective. As seen there, the two years before 1981 include some of the lowest (although not the very lowest) values of the monthly ratio. However, similar intervals of very low ratios occur also at other times, both before and after this two-year period. The 1981-1982 interval of above-average monthly ratios is even less exceptional. The right panel of Fig. 3 shows several similar short (1-2-year) intervals (humps) with even larger values of monthly ratios than those in 1981-1982. At many humps there is a large increase between pre-hump monthly ratios and during-hump ratios in the same way as around 1980/1981. Moreover, since the typical length of such humps seems to be about 1-2 years, a choice of two years maximizes the difference of the two successive short intervals. If one would select a bit longer period than two years for such a comparison, the corresponding means would be much closer to the long-term averages.

We note that these humps, like the 1980 jump, mostly occur at solar maxima, for example, around 1990, 2001 and, especially, with the highest ratios around 2014. A similar connection was found earlier by Tapping & Valdés (2011) and Tapping & Morgan (2017) for both the original (version 1) and the revised (version 2) sunspot number series. The 1981-1982 interval is the first solar maximum where such humps of large ratios appeared. While none of these humps lead to a permanent increase in the ratios, they all contribute to its long-term increase.

We also included in the right panel of Fig. 3 two running mean curves of the ratio with a roughly 2-year (25-month; black curve) and a roughly 10-year (121-month; red curve) length. The black curve shows the above discussed humps clearly. While such humps existed even before 1980, their amplitude increased considerably since the 1981-1982 hump. Accordingly, the claim by Clette (2021), although erroneous, is quite understandable. The red curve shows the long-term evolution of the F10.7-sunspot ratio. It shows that, while there was a large increase from the 1970s to 1980s due to the 1980-1981 hump, the ratio did not reach a plateau but, rather, continued increasing until the 2010s.

As the above analysis shows, selecting two 2-year intervals around the maximum of cycle 21 indeed leads to a large difference between the two 2-year intervals. However, this 1980 jump does not represent a permanent level increase in these ratios. Similar pairs of short intervals where the ratio increases momentarily can be found around all later solar maxima. So, the 1980 jump is just the first of a series of short-term humps that mostly occur during solar maxima. None of these humps lead to a permanent increase in the F10.7/sunspot ratio, but all of them contribute to that. Rather than a step-like change, the right panel of Fig. 3 suggests a more systematic increase in the ratio of the F10.7 index and the correlated sunspot number. We will study next the long-term evolution of this ratio more generally using, in addition to the 10.7 cm flux measured in Canada, four other radio fluxes (F30, F15, F8, and F3.2) measured completely independently in Japan.

## 5. Sunspot - radio flux relations

We have correlated the yearly mean sunspot numbers with each of the five solar radio fluxes separately. The corresponding five scatterplots are depicted in the five panels of the left column of Fig. 4, together with their best-fit lines. (We will discuss the small nonlinearity in the relations between the different parameters in Sec. 8.5. It turns out that there is hardly any difference between results obtained from a linear model or a nonlinear model). Correlation coefficients and p-values of correlation, as well as the slopes and intercepts of the best-fit lines are given in Table 1. One can see in Table 1 that all five correlation coefficients are very large, with statistical significance beyond any uncertainty. This reflects the above mentioned similarity of the five radio flux time series and the sunspot number.

The correlation coefficient between sunspots and F10.7 is slightly larger than for other radio fluxes and the corresponding p-value is somewhat smaller for F10.7 than for other fluxes. The p-value also depends on the length of the time series, and it is reassuring to see that the two longest radio fluxes (F10.7 and F8) yield the smallest p-values. Also, the largest p-values (still very small) are found for the least correlating F3.2 flux and for the shortest time series of the F15 flux. Moreover, the slope of the F10.7 best-fit line is clearly larger than the slope of any other radio flux. This shows that the 10.7 cm radio flux is indeed the most sensitive out of the five radio wavelengths to measure the type of solar magnetic activity that is quantified by the sunspot number.

The five panels of the right column of Fig. 4 show the yearly differences (residuals) between each radio flux and the corresponding scaled sunspot number normalized by the yearly radio flux. We have also included the best-fit lines for these differences and also the lines with slopes that are two standard deviations (2 $\sigma$) above or below the slope of the best-fit line. We have depicted these three slopes for all five cases in Table 2.

Figure 4 and Table 2 show that all five radio fluxes increase with respect to the sunspot number during this time interval. The long-term and even the inter-annual evolution of the five differences are reasonably similar, although some systematic differences are also seen. The two longest-wavelength fluxes (F30 and F15) have their minima in the 1960s, the two shorter-wavelength fluxes (F10.7 and F8) have roughly equal minima in the 1960s and 1970s, and the shortest-wavelength flux (F3.2) in the late 1970s. There are also some differences in the location of maxima between the five differences. While the (normalized) difference maxima of the four longer-wavelength fluxes are in the 2010s, for the shortest-wavelength it was already in the early 2000s.

For all five fluxes, a permanent turn, as defined by the best-fit line, from the negative difference (relatively stronger sunspot number) to positive difference (relatively stronger radio flux) takes place in the time interval from the early 1980s to early 1990s. All differences depict a hump in 1981-1982, most clearly seen in the three middle wavelengths (F15, F10.7, and F8), where the differences rise above zero. This hump indicates a common, momentary enhancement in the changing relation between the radio fluxes and the sunspot number. This hump also explains the above discussed 1980 jump (Clette 2021), and shows that it is a physical phenomenon rather than a problem in the calibration or (post)processing of the (Canadian) 10.7 cm measurements. Indeed, the great similarity in the evolution of the difference related to the F10.7 flux measured in Canada and the differences of the four other radio fluxes measured in Japan simply rule out the possibility of the 1980 jump or any of the changes depicted in Fig. 4 being due to some speculative problem in F10.7. Also, rather than a single jump from one level to another, Figure 4 shows that the change across 1980/1981 was only one in a series of several temporary increases (humps) embedded in a long increasing trend in the relation between all radio fluxes and the sunspot number.

**Table 1.** Correlation coefficients, p-values, and the slopes and intercepts of the best-fit lines between yearly means of the five radio fluxes and the sunspot number depicted in the left panel of Fig. 4.

|       | CC    | p       | Slope | Intercept |
|-------|-------|---------|-------|-----------|
| F30   | 0.982 | 4.0e-47 | 0.412 | 41.65     |
| F15   | 0.988 | 2.0e-49 | 0.562 | 47.46     |
| F10.7 | 0.993 | 1.7e-65 | 0.650 | 62.56     |
| F8    | 0.991 | 3.0e-60 | 0.546 | 72.25     |
| F3.2  | 0.978 | 2.4e-44 | 0.434 | 253.59    |

Accordingly, Figure 4 strongly suggests that a real (physical), systematic and quite a large change is going on in the Sun from the late 1970s until 2010s. The increase in the (normalized) difference in Fig. 4 is largest and most significant in the longest-wavelength flux F30, as indicated by the largest slope of the F30 best-fit line and the smallest p-value (see Table 2). Interestingly, the slopes decrease with the decreasing wavelength almost systematically. This suggests that there is a temporal change proceeding in the Sun during the studied time interval which affects at least a large fraction of solar spectrum at radio frequencies.

We note that, although we have analyzed the (normalized) differences in Fig. 4 in terms of a linear model, their temporal evolution is hardly purely linear or monotonous in time. In fact, most differences in Fig. 4 depict some evidence for oscillatory behavour. This is particularly true for the two radio fluxes (F10.7 and F8) with the longest temporal extent. These two differences in Fig. 4 decrease, rather than increase, during their first decade from the 1950s to 1960s. The two other differences (F30 and F15) also depict a decrease during their first decade. Moreover, most differences depict a decrease from a maximum in the 2010s (or 2000s for F3.2) until the end of data. Accordingly, there is evidence that the relation between radio fluxes and the sunspot number is oscillating at the half-period of about 40-50 years. The period of increase discussed here also lasted some 40-50 years from the 1960s-1970s to 2010s, that is, during the decay of the Modern Maximum. Accordingly, it is quite likely then that the long-term oscillation in the relation between radio fluxes and the sunspot number is related to the roughly centennial Gleissberg cyclicity.

## 6. Mutual relation of radio fluxes

In order to study the mutual relation between the different radio fluxes in more detail, we have correlated the yearly means of the F10.7 flux with each of the other four radio fluxes separately. The corresponding four scatterplots are depicted in the four panels of the left column of Fig. 5, together with their best-fit lines. The correlation coefficients and p-values of correlation, as well as the slopes and intercepts of the best-fit lines are given in Table 3.

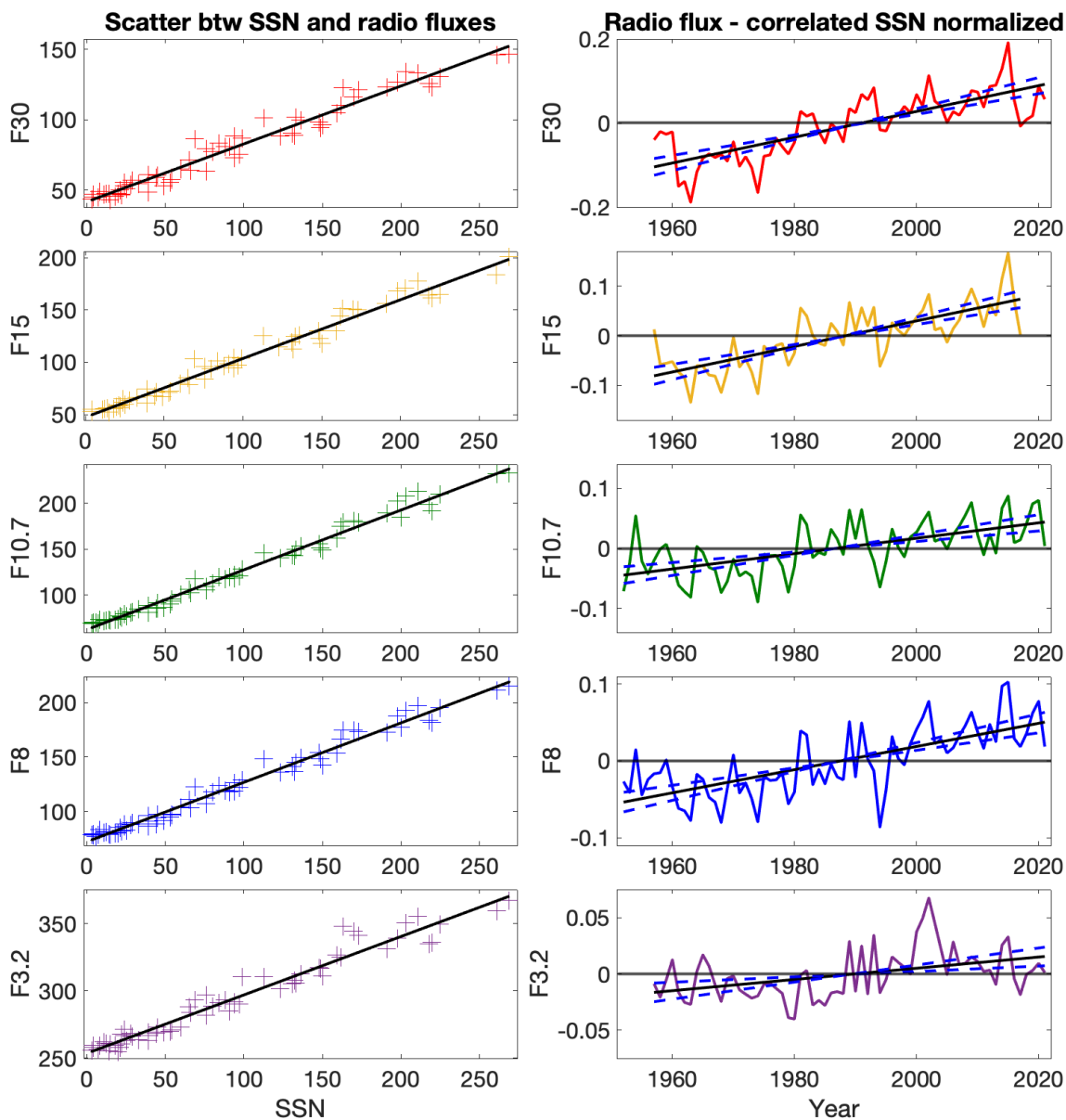

**Fig. 4.** Correlation between the yearly sunspot number and the yearly means of the five radio fluxes. Left column: Scatterplots between sunspot number and radio fluxes (colored crosses), together with corresponding best-fit lines (black). Right column: Differences between the yearly radio fluxes and the correlated sunspot number normalized by the corresponding radio flux. Best-fit lines (black solid lines) and lines with slopes that are two standard deviations above or below the best-fit line slope (two blue dashed lines) are also included. Order of radio wavelengths and their colors are the same as in Fig. 1.

**Table 2.** Correlation coefficients, p-values, and the slopes of the best-fit lines for the yearly normalized differences between the five radio fluxes and the correlated sunspot number depicted in the right panel of Fig. 4, together their $2\sigma$ upper and lower values. All slope values are multiplied by 1000.

|       | CC    | p       | Slope | +2σ slope | -2σ slope |
|-------|-------|---------|-------|-----------|-----------|
| F30   | 0.779 | 2.2e-14 | 3.068 | 3.690     | 2.446     |
| F15   | 0.762 | 9.6e-13 | 2.573 | 3.142     | 2.004     |
| F10.7 | 0.609 | 2.3e-08 | 1.281 | 1.686     | 0.876     |
| F8    | 0.703 | 1.2e-11 | 1.51  | 1.881     | 1.140     |
| F3.2  | 0.436 | 2.9e-04 | 0.498 | 0.757     | 0.239     |

One can see in Table 3 that each of the four Japanese radio fluxes correlate even better with the F10.7 flux than with the sunspot number. This is very reassuring and supports the long-term homogeneity and stability of all five radio fluxes. It is also reassuring to note that the fluxes with wavelengths closest to F10.7 (F8 and F15) have the largest correlation coefficients and slopes and the smallest p-values for correlation with F10.7. These results further underline the long-term homogeneity and stability of the Canadian 10.7 cm flux, and give additional evidence that the change around 1980 is not due to an error in F10.7 measurements.

We also note that the intercept of F30 and F15 is very small and that of F8 fairly small, indicating that these four radio fluxes (F30, F15, F10.7, and F8) are practically multiplicative of each other, at least at the yearly resolution. This indicates that the respective source regions respond very similarly to the long-term changes that proceed in the Sun. On the other hand, the inter-cept of the F3.2 flux, which is located further away in the solar spectrum from the four other radio frequencies, is much larger, suggesting that the respective source region responds quite differently, with a much smaller relative increase to the long-term change in the Sun.

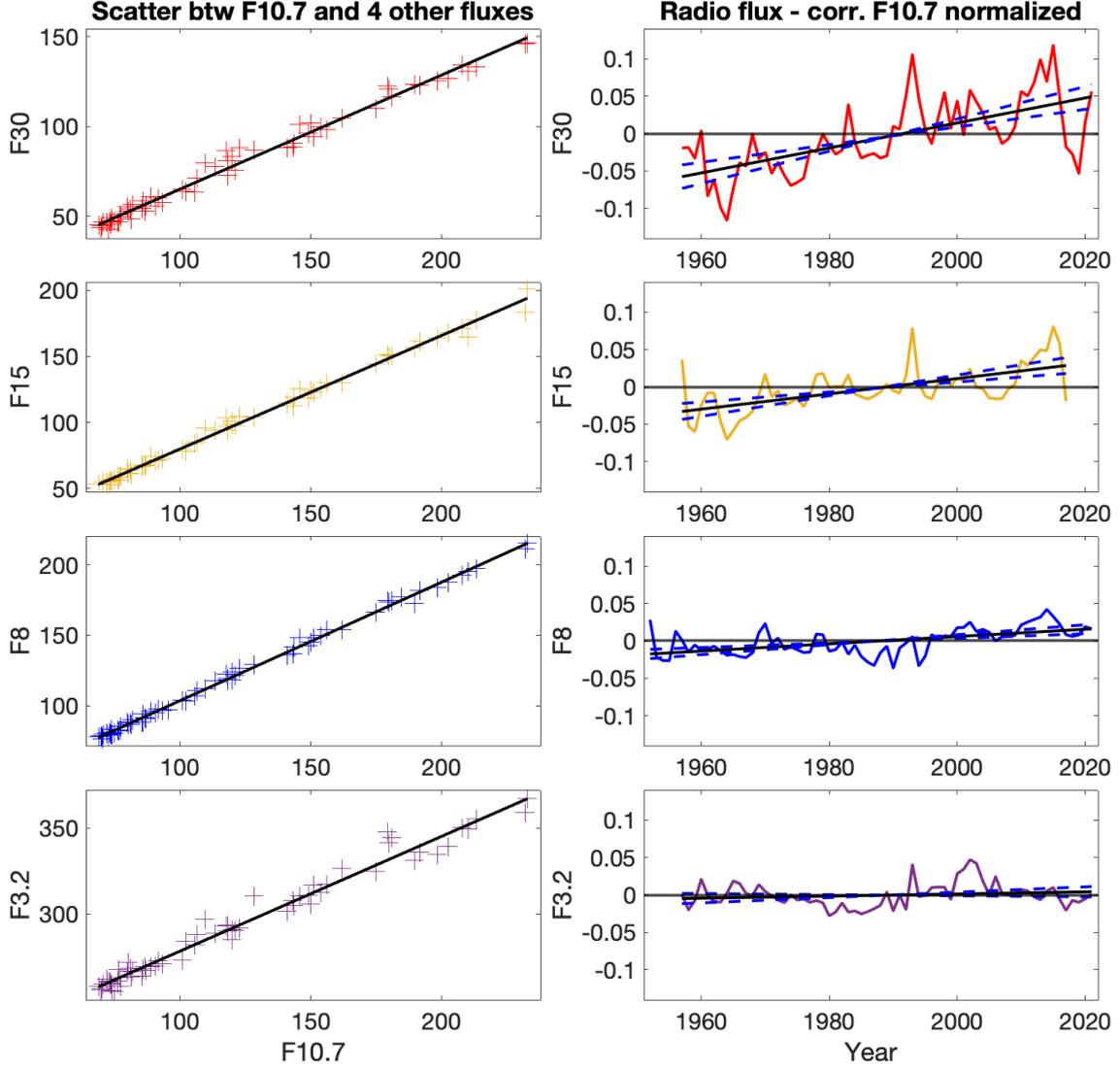

**Fig. 5.** Correlation between the yearly values of the 10.7 cm radio flux and the four other radio fluxes. Left column: Scatterplots between 10.7 cm and the four radio fluxes (colored crosses), together with the corresponding best-fit lines (black). Right column: Differences between the four radio fluxes and the correlated 10.7 cm radio flux normalized by the corresponding radio flux. Best-fit lines (black solid lines) and lines with slopes that are two standard deviations above or below the best-fit line slope (two blue dashed lines) are also included. Order of radio wavelengths and their colors are the same as in Fig. 2. (F10.7 is naturally omitted.)

**Table 3.** Correlation coefficients, p-values, and the slopes and intercepts of the best-fit lines between yearly means of the F10.7 cm radio flux and the four other radio fluxes depicted in the left panel of Fig. 5.

|      | CC    | p       | Slope | Intercept |
|------|-------|---------|-------|-----------|
| F30  | 0.993 | 7.0e-60 | 0.634 | 1.80      |
| F15  | 0.997 | 1.7e-66 | 0.858 | -5.80     |
| F8   | 0.998 | 4.4e-87 | 0.840 | 19.61     |
| F3.2 | 0.988 | 2.5e-52 | 0.668 | 211.68    |

**Table 4.** Correlation coefficients, p-values, and the slopes of the best-fit lines for yearly normalized differences between the four Japanese radio fluxes and the correlated F10.7 radio flux depicted in the right panel of Fig. 5, together their $2\sigma$ upper and lower values. All slope values are multiplied by 1000.

|      | CC    | p       | Slope | +$2\sigma$ slope | -$2\sigma$ slope |
|------|-------|---------|-------|------------------|------------------|
| F30  | 0.651 | 4.4e-09 | 1.669 | 2.159            | 1.178            |
| F15  | 0.600 | 3.2e-07 | 1.022 | 1.376            | 0.668            |
| F8   | 0.549 | 8.8e-07 | 0.486 | 0.666            | 0.306            |
| F3.2 | 0.165 | 0.19    | 0.142 | 0.357            | -0.072           |

The panels in the right column of Fig. 5 show yearly differences (residuals) between the four other radio fluxes and the correlated F10.7 flux normalized by the respective yearly radio fluxes. We have also included in these panels the best-fit lines for the (normalized) differences and the lines with slopes that are two standard deviations ($2\sigma$) above or below the best-fit slope. The correlation coefficients, p-values and the three slopes for the four differences are given in Table 4.

Figure 5 and Table 4 show that the three long-wavelength fluxes (F30, F15, and F8) increase with respect to F10.7 during this time interval statistically significantly (very small p-values). In the case of the two longest wavelengths (F30, and F15), this increase is quite large (slopes of $1.669 \cdot 10^{-3}$ and $1.022 \cdot 10^{-3}$) and very significant (p-values less than $3.2 \cdot 10^{-7}$). The increase is largest in F30 (about 11%; see later in Section 8) and quite considerable in F15 (about 6%), even over a slightly shorter interval. Although the increase for F8 is statistically significant, it is quite small (about 3%) and somewhat less significant than for F30 and F15. For F3.2 there is no statistically significant linear change with r espect to F10.7 over this time interval.

Interestingly, the fact that we find statistically significant long-term changes also between the different radio fluxes gives evidence that the change in the relation between sunspot number and the five radio fluxes depicted in Fig. 4 is not due, for example, to a significant and systematic error in the sunspot number (either), or due to a possible small nonlinearity in the relation between sunspot and the radio fluxes. Rather, these results give strong additional evidence for an important, so-far little understood systematic long-term change in the Sun, which affects the sources (possibly the mean source heights; see later in Section 8) of the different radio emissions and the different types of magnetic structures differently in a consistent manner.

Finally, we note that the temporal evolution of the relation between F10.7 and the two longer-wavelength fluxes (F30 and F15) depicted in Fig. 5 is mutually very similar, and quite similar even to the relation between these two fluxes and the sunspot number (see Fig. 4). This gives additional evidence for the physical validity of these changes. Moreover, the similarity of these differences in the two Figures emphasizes the above discussed possibility of a longer-term, oscillatory variation in the Sun related to the Gleissberg cycle.

## 7. Other solar parameters

In order to have an even more complete view of the different facets of this surprising long-term change in the Sun we have also studied here two other, rather different chromospheric parameters. Although the F10.7 flux is often used as a proxy for solar UV/EUV irradiance, especially for long time scales, we will also use here the core to wing ratio of the singly ionized magnesium (MgII), which is a standard proxy for solar UV/EUV irradiance and whose intensity is closely related to magnetic field strength. On the other hand, the number of active regions takes into account those regions that are magnetically active but have a considerably weaker field intensity than sunspots. The number of sunspots and the number of active regions together cover a fairly large range of magnetically active regions on solar surface.

The scatterplots between the yearly MgII index and the sunspot number, and between the yearly number of active regions and the sunspot number are depicted in the left column of Fig. 6, together with their best-fit lines. The correlation coefficients and p-values of correlation, as well as the slopes and intercepts of the best-fit lines for these two cases are given in Table 5. One can see in Table 5 that both correlations are very good and highly statistically significant, with MgII having roughly the same level of correlation with the sunspot number as the radio fluxes have (see Table 1), and the correlation of active regions with sunspots remaining only slightly smaller. We also note that the p-values in Table 5 cannot be compared with each other or with those given in the earlier corresponding Tables because of a much shorter time interval of available data in Fig. 6.

**Table 5.** Correlation coefficients, p-values, and the slopes and intercepts of the best-fit lines between the yearly sunspot number and the yearly means of two other solar parameters (MgII index and number of active regions) depicted in the left panel of Fig. 6.

|       | CC    | p        | Slope     | Intercept |
|-------|-------|----------|-----------|-----------|
| **MgII** | 0.988 | 4.81e-35 | 8.20e-05  | 0.15      |
| **AR**   | 0.971 | 3.14e-25 | 0.72      | 11.58     |

The panels in the right column of Fig. 6 show the yearly differences between the MgII index and the correlated sunspot number (top) and between the number of active regions and the correlated sunspot number (bottom). We note that these differences are not normalized by the corresponding yearly values unlike in Figs. 4 or 5, because the small number of active regions in minimum years would distort the ratio and give an erroneous view. As in the other similar figures, we have also included in these two panels the best-fit lines for the differences and the lines with slopes that are two standard deviations ($2\sigma$) above or below the best-fit slope. The correlation coefficients, p-values and the three slopes for the two differences are given in Table 6.

Figure 6 and Table 6 show that both the MgII index and the number of active regions increase with respect to the sunspot number during the respective time intervals. In both cases the increase is statistically significant (p-values less than $2.5 \cdot 10^{-3}$), and the slopes deviate from zero at the level of significance which is well beyond $2\sigma$. However, because of the shorter time interval (and the slightly smaller correlation coefficient in the case of ARs), the significance of correlation (p-value) is weaker than in the case of correlation of radio fluxes with the sunspot number.

The relative increase in the MgII index during the 43 years is only about 0.9% (see Section 8). However, as mentioned above, the solar cycle variation of the MgII index is also rather small, only about 6%. Accordingly, the increasing trend forms a considerable fraction (roughly 15%) of the solar cycle variation of the index. For the number of active regions, the relative increase is much stronger, about 25%. Interestingly, however, the increasing trend of the active regions forms almost the same fraction, about 16% of the solar cycle variation as in the case of the MgII index.

Finally, we note that, since the two parameters depicted in Fig. 6 do not cover much of the pre-1980 time interval, their (relative) increase cannot be straightforwardly compared with the increase in radio fluxes over the longer time interval. However, the results in Fig. 6 and Table 6 show that even other solar parameters than radio fluxes do experience a significant increase with respect to the sunspot number during the last several decades. We also note that the detailed short-term variations in the difference between MgII and the correlated sunspot number (upper right panel in Fig. 6) are quite similar to those in the corresponding differences of radio fluxes. In particular, the hump in 1981-1982 is clearly visible, as well as the peaks in the early 2000s and mid-2010s that are also seen in the corresponding differences of the radio fluxes. However, the short-term variations in the differences between the number of active regions and the sunspot number do not have much resemblance with those in the other differences.

## 8. Discussion

### 8.1. No inhomogeneity in F10.7

We have shown here that not only the Penticton 10.7 cm radio flux but also all four Japanese radio fluxes depict a significant relative increase with respect to the sunspot number. This result alone rejects the hypothesis (Clette 2021) of a permanent step-like rise (1980 jump) in the mutual relation between the 10.7 cm radio flux and the sunspot number due to an inhomogeneity in the 10.7 cm radio flux. Rather than a permanent jump, there is a roughly 2-year local maximum (hump) in this relation in 1981-1982, which is the first in a series of similar local maxima to occur later, most of them located close to solar maxima and being higher than the hump in 1981-1982.

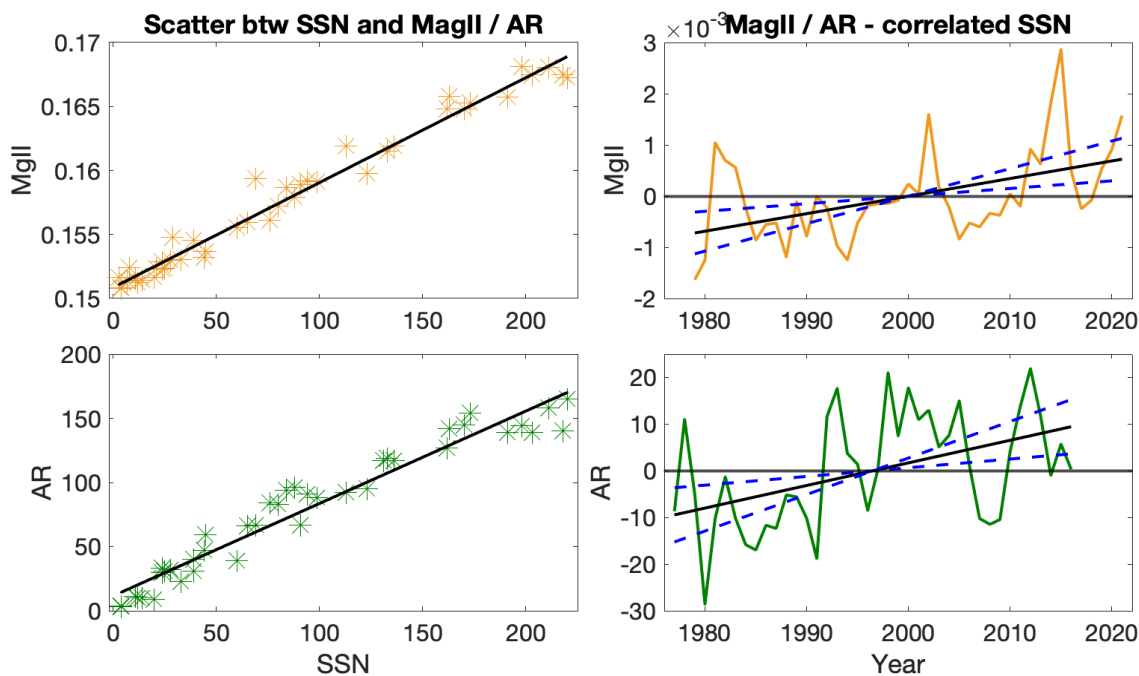

**Fig. 6.** Correlation between the yearly values of two solar parameters and the sunspot number. Left top: Scatterplot between the MgII index and the sunspot number (colored asterisks), together with the corresponding best-fit lines (black). Left bottom: Scatterplot between the yearly number of active regions and the sunspot number. Right top: Difference between the MgII index and the correlated sunspot number normalized by the MgII index. Best-fit line (black) and lines with slopes that are two standard deviations above or below the best-fit line slope (two blue dashed lines) are also included. Right bottom: Corresponding normalized difference between the yearly number of active regions and sunspots.

**Table 6.** Correlation coefficients, p-values, and the slopes of the best-fit lines for yearly differences between the MgII index (upper row) or number of active regions (lower row) and the correlated sunspot number depicted in the right panel of Fig. 6, together their $2\sigma$ upper and lower values. All slope values are multiplied by 1000.

|      | CC    | p        | Slope  | $+2\sigma$ slope | $-2\sigma$ slope |
|------|-------|----------|--------|------------------|------------------|
| MgII | 0.483 | 1.04e-03 | 0.034  | 0.0540           | 0.015            |
| AR   | 0.467 | 2.41e-03 | 484.37 | 782.31           | 186.43           |

Clette (2021) conclude in italic font that (their study) "..allows us to confirm that the 1980 jump definitely occurs in the F10.7 series and is not due to any unsuspected and uncorrected flaw in the SN series." We have shown here that this conclusion is erroneous and that there is no inhomogeneity in the F10.7 series. This is also in agreement with the analysis by Tapping & Morgan (2017), who did not find the suggested abrupt increase. Indeed, the 10.7 cm radio flux is homogeneous, maybe the very most reliable and homogeneous index of solar activity of all parameters. However, we do not find any error in the sunspot number either. The change seen in the mutual relation of the many solar parameters included in this study is inherent to the Sun: solar magnetic fields are changing. This change is large and affects the different solar parameters in a slightly different way, suggesting that the conditions in the respective layers of the solar atmosphere are modified slightly differently.

### 8.2. Results for radio fluxes

The relative increase in each of the five radio fluxes with respect to the sunspot number is depicted in Table 7. As mentioned above, the increase percentages almost systematically increase with the wavelength of the radio emission. The increase in the F30 flux, about 20%, is very large, especially as most of it comes from only a part of the whole time interval. Without the first and last 5-year sections of the time interval covered, the increase would be even larger. As noted earlier, the evolution of the relation between radio fluxes and sunspot number during the first and last decades may in fact indicate (see Fig. 4) that this relation is not linear in time but oscillates at the centennial Gleissberg cycle.

The effect of different first years affects to a small difference in the increase percentages between the two radio fluxes (F10.7 and F8) with the longest time interval and the three other fluxes with a shorter interval. Taking these two longer fluxes to start in 1957 would yield slightly larger increase percentages of about 9.44% for F10.7 and 11.09% for F8. Accordingly, the small difference in the temporal extent of the five radio fluxes does not much affect the conclusions made based on the percentages given in Table 7.

The fact that the two longest radio wavelengths experience the largest relative increase with respect to the sunspot number also leads to the fact that they increase with respect to the F10.7 flux. The respective increase percentages are depicted in Table 8 for those three fluxes for which this increase was found to be significant. As discussed above, for F3.2 there is no significant change and even for F8 the change is quite small. However, the F30 and F15 fluxes increase with respect to the F10.7 flux by a considerable fraction. Since the longer radio waves come, on average, from higher altitudes in the solar atmosphere than the shorter radio waves, this indicates an interesting structural change in the solar atmosphere.

**Table 7.** Slopes of the best-fit lines for yearly differences between the five radio fluxes and the correlated sunspot number, start year, end year, and relative increase over the whole time extent of each flux as a percentage. All slope values are multiplied by 1000.

|       | Slope | Start year | End year | Increase% |
|-------|-------|------------|----------|-----------|
| F30   | 3.068 | 1957       | 2021     | 19.94     |
| F15   | 2.573 | 1957       | 2017     | 15.70     |
| F10.7 | 1.281 | 1952       | 2021     | 8.97      |
| F8    | 1.51  | 1952       | 2021     | 10.57     |
| F3.2  | 0.498 | 1957       | 2021     | 3.24      |

**Table 8.** Slopes of the best-fit lines for yearly differences between the three longer radio fluxes and the correlated F10.7 flux, start year, end year, and relative increase over the whole time extent of each flux as a percentage. All slope values are multiplied by 1000.

|     | Slope | Start year | End year | Increase% |
|-----|-------|------------|----------|-----------|
| F30 | 1.669 | 1957       | 2021     | 10.85     |
| F15 | 1.022 | 1957       | 2017     | 6.23      |
| F8  | 0.486 | 1952       | 2021     | 3.4       |

### 8.3. Consequences to ionosphere and atmosphere

Comparing F10.7 and F30 radio fluxes to the ionospheric critical height (electron density) observations at several stations, Laštovička & Burešová (2023) could see a difference in the response of the ionosphere to these two radio fluxes over the time interval from 1976-2014. The ionospheric response was relatively larger than predicted either by F10.7 or sunspots, but was in an agreement with F30. As a consequence, Laštovička & Burešová (2023) suggested that F30 should be used in the future to describe the ionospheric response, rather than the conventionally used F10.7. These results are in an excellent agreement with the current results on different long-term evolutions in sunspots, F10.7 and F30, and verify the validity of the larger long-term increase in F30 than F10.7 (or sunspots). They also show that there are practical consequences from this difference in the temporal evolution of the different radio fluxes. On the other hand, our analysis of the five radio fluxes verifies that the results obtained by Laštovička & Burešová (2023) are not caused by the suggested, hypothetical inhomogeneity in F10.7.

As another piece of evidence for such practical consequences, we note that it has been demonstrated that the DTM-2013 model (Drag Temperature Model) for the atmospheric drag of satellites (Bruinsma 2015) performs best in recent years with the F30 radio flux at altitudes lower than 500 km (Dudok de Wit et al. 2014).

### 8.4. Results for MgII index and active regions

We found that also the two other solar parameters studied here, the MgII index and the number of active regions, depict a significant long-term increase with respect to the sunspot number. Table 9 depicts the corresponding increase percentages. Although the increase in the MgII index with respect to the sunspot number is significant, its increase percentage is only 0.93% during the last 40 years. On the other hand, the number of active regions increased relatively more with respect to the sunspot number: the increase was nearly 25% from 1977 to 2016.

**Table 9.** Slopes of the best-fit lines for yearly differences between the MgII index and the number of active regions (ARs) with the correlated sunspot number, start year, end year, and relative increase over the whole time extent as a percentage. All slope values are multiplied by 1000.

|      | Slope   | Start start yearyear | End year | Increase% |
|------|---------|----------------------|----------|-----------|
| MgII | 0.034   | 1979                 | 2021     | 0.93      |
| AR   | 484.372 | 1977                 | 2016     | 24.67     |

Despite these different percentages, it is interesting to note that the increasing trend makes roughly an equal fraction (about 15%) of the solar cycle amplitude for each of the two parameters. Taking into account that both the MgII index and the active regions used here are chromospheric parameters, this similarity is hardly a coincidence but, rather, reflects the typical level of relative change in the solar atmosphere at chromospheric altitudes.

It is interesting to note that the number of active regions has no statistically significant trend with respect to the F10.7 flux. This is depicted in Fig. 7. The coefficient for their correlation is moderate (cc= 0.242) and indicates some increase. However, the p-value of this correlation is 0.133, indicating that there is no statistically significant correlation. Accordingly, the F10.7 flux and the number of active regions roughly follow each other in the changes taking place in the Sun during the decay of the MM.

This is very reassuring, taking into account that the variable part of F10.7 emissions mainly (up to more than 60%) comes by bremsstrahlung mechanism from the non-sunspot active regions, while only 10% of F10.7 flux come from sunspots generated mainly by gyroresonance mechanism (Schonfeld et al. 2015). The small positive slope of the AR-F10.7 difference (see the right panel of Fig. 7), although not statistically significant, is in line with the view that, due to the small contribution from sunspots, the F10.7 flux increased somewhat less with respect to the sunspot number than the number of active regions.

### 8.5. Effect of nonlinearity

It is known since long that the daily sunspot number and the contemporaneous daily F10.7 index do not have a purely linear relation but, rather, their relation is slightly nonlinear both at the smallest and the largest values (see, e.g., Tapping & Morgan 2017; Clette 2021). Similar nonlinearities have been found in the daily values between practically all photospheric and chromospheric parameters (see, e.g., Foukal 1998; Yeo et al. 2020). The reason for these nonlinear relations between daily values have been suggested to be due to the fact that the lifetime of chromospheric plages (and, thus, active regions) is longer than sunspots (see, e.g., Preminger & Walton 2007). The effect of the longer lifetime of plages was studied by Preminger & Walton (2005, 2007) and later by Yeo et al. (2020) using finite-impulse response filtering, which led to a nonlinear relation and an improved correlation between the contemporaneous daily values of sunspots and plages (more generally, between photospheric and chromospheric parameters).

Supposing that the yearly values of sunspots and radio fluxes would be strongly nonlinear, with radio fluxes saturating with high sunspot numbers, the weakening solar activity would raise the radio flux with respect the sunspots and, thereby, produce the observed difference in the long-term evolution of sunspots and radio fluxes. In order to test the viability of this explanation, we have studied the nonlinear relation of the yearly values of sunspots and two radio fluxes (F10.7 and F30). The left panels of Fig. 8 show the best-fit lines (repeated from Figs. 4 and 5) and the best-fit higher-degree (third or second; blue curve) polynomials between the three parameters. In the case of the relation between sunspots and the two radio fluxes, the best-fitting polynomial is of third order, in the case of the relation between F30 and F107 it is of second order. For these polynomials the coefficient of the highest power of the nonlinear fit is significantly different from zero and the adjusted correlation coefficient reaches its maximum. The nonlinearity is very small (although significant) and visible in Fig. 8 only at the highest yearly values.

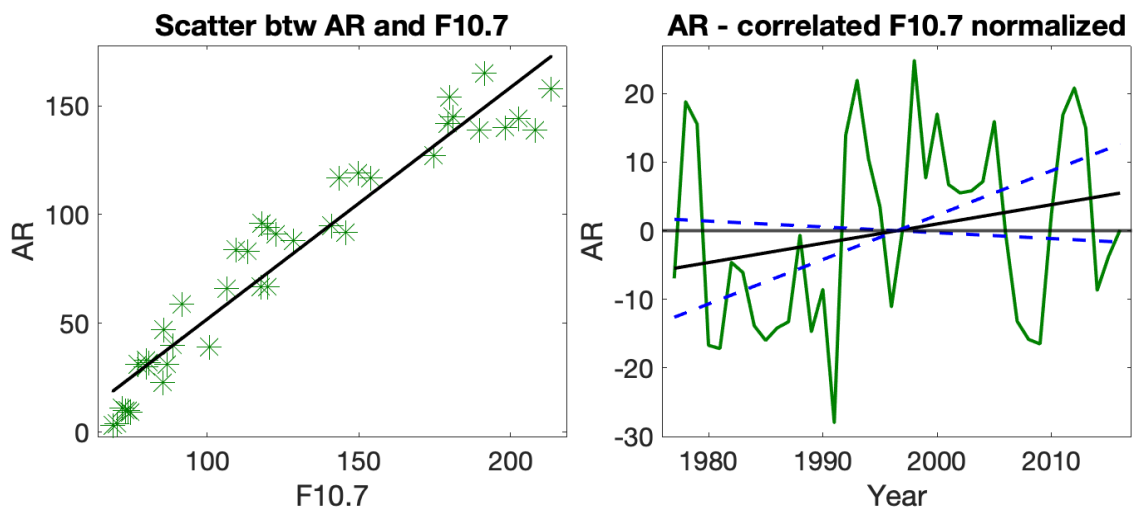

**Fig. 7.** Correlation between the yearly number of active regions and the F10.7 flux. Left: Scatterplot between the active regions and the F10.7 flux (green asterisks), together with the corresponding best-fit line (black). Right: Difference between the yearly number of active regions and the correlated F10.7 flux. Best-fit line (black) and lines with slopes that are two standard deviations above or below the best-fit line slope (two blue dashed lines) are also included.

The right panels of Fig. 8 depict the related normalized differences (residuals) for both linear and nonlinear fits between the two respective parameters of the left panels. The right panels of Fig. 8 show that the temporal evolutions of the normalized differences of the linear and nonlinear fits follow each other very closely in each of the three cases. (The same result is, naturally, obtained also for unnormalized differences; not shown). The linear and nonlinear differences are in a remarkably good agreement for sunspots and F30 (top right panel of Fig. 8) where they visibly deviate from each other only in few years. For the two other cases, the linear and nonlinear differences agree slightly less perfectly, and there is a visible difference between them in almost every year. In the relation of sunspot number and F107 the linear fit produces clearly larger differences in 5-6 years after 2000, while the nonlinear fit is larger only in 3-4 years at the same time. This makes the nonlinear difference rise slightly more slowly than the linear fit. In the F30-F107 relation, the linear fit gives slightly lower minima and higher maxima than the nonlinear fit, but the trends are very similar. Concluding, Figure 8 shows convincingly that the temporally changing relation between the yearly values of the radio fluxes and the sunspot number is not due to the small nonlinearity in their relationship. (We selected only the three most important relations in Fig. , but the same conclusion was verified to be valid for all relations).

The notable difference in the relation of sunspots and plages (in general photospheric and chromospheric parameters) when using either daily (considerably nonlinear) or yearly (weakly nonlinear) values may first seem rather surprising. Foukal (1998) showed clearly how the daily values of sunspot and plage areas have a highly nonlinear relation, while the yearly means of the same parameters were found to have an almost purely linear relation, in an analogy to what was found here. They also constructed a simple model which could explain this difference in terms of a different dependence of sunspot and plage lifetimes on their areas. Plage lifetimes increase with plage area even for the largest plages, but sunspot lifetimes do not. This compensates the decreasing plage-to-spot area ratio with increasing area, which follows from to the nonlinear relation of daily values. This explanation is likely to apply to the current situation as well, since sunspot number and sunspot area correlate well and radio fluxes are dependent on the area of active regions (plages).

### 8.6. Notes on earlier studies

There are not many studies in recent decades that examine the long-term evolution of solar radio fluxes at different wavelengths over several solar cycles. In one early study, of which we became aware only after the completion of this paper, Nicolet & Bossy (1985) used radio fluxes at the same five wavelengths in 1957 - 1983 as used here, verifying that each of these wavelengths varied closely in phase with the sunspot cycle, but with different cycle amplitudes. Using yearly means of these fluxes they found that they all correlated linearly with each other within an accuracy of about ±2-5%. This agrees very well with our finding of dominantly linear nature of the mutual relation of these fluxes. They also found that the relation between daily fluxes of the longer wavelengths, for example, between F30 and F107, was somewhat nonlinear and was best fit with a quadratic relation, similarly as we found for the small nonlinearity of yearly means for these fluxes (see the bottom panel of Fig. 8). Although no proper temporal analysis was made in this paper, some correlation plots show that longer wavelength fluxes were relatively slightly lower in the 1960s and higher in the 1970s, which is in agreement with our results. However these differences in Nicolet & Bossy (1985) were all within the estimated few-percent accuracy of respective correlations, and no claim on temporal changes in correlations were made in Nicolet & Bossy (1985). This may be due to the fact that the analysis of this paper did not yet include the time interval of the largest change.

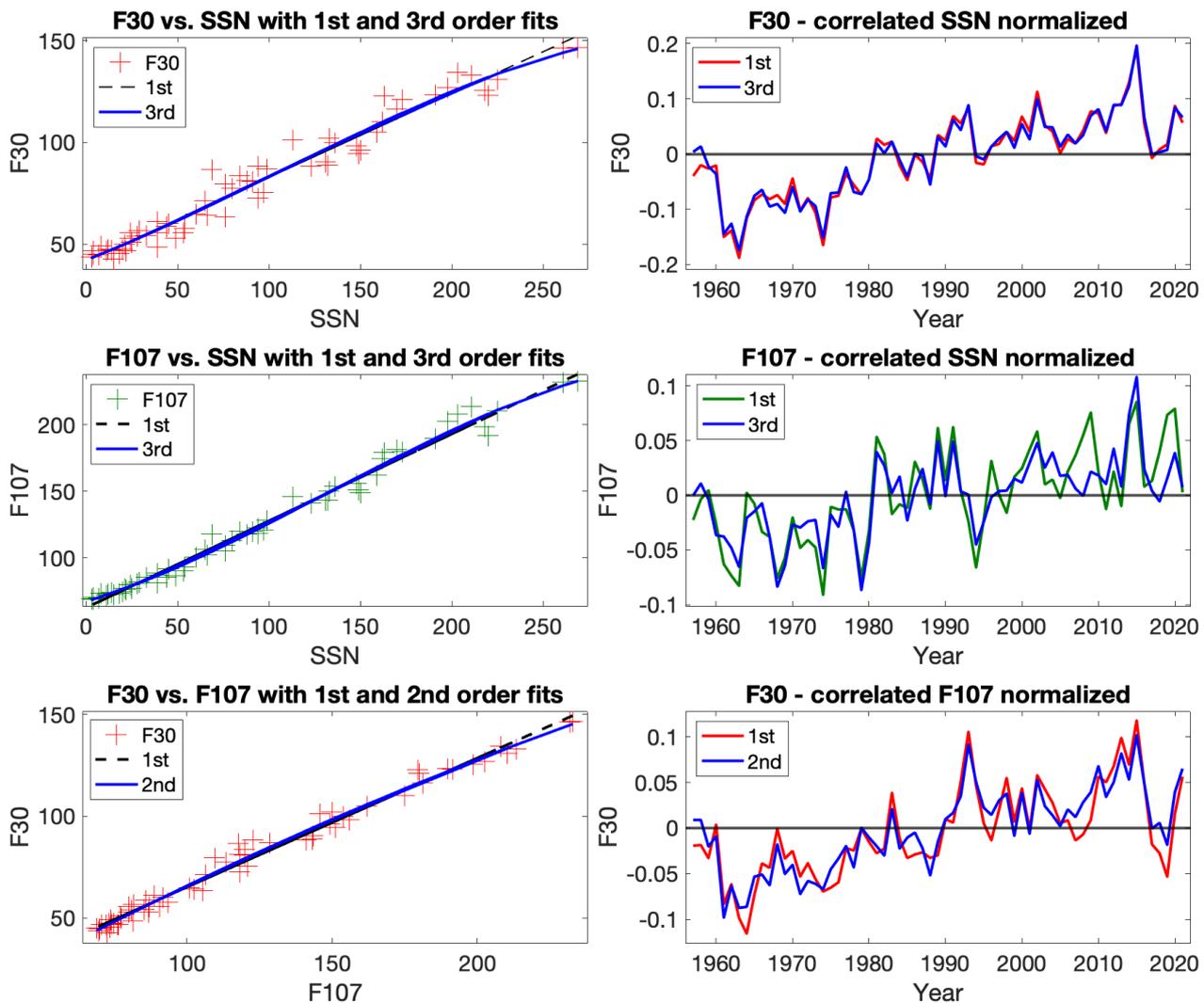

**Fig. 8.** Correlation between the yearly values of two radio fluxes and the sunspot number with linear and nonlinear fits. Left: Scatterplots between the F30 flux and the sunspot number (top row; red asterisks), the F107 flux and the sunspot number (middle row; green asterisks) and the F30 and F107 fluxes (bottom row; red asterisks), together with the corresponding best-fit line (black dashed line) and best-fit polynomial (blue curve). Right: Normalized differences between the two parameters in the corresponding left panel using their linear (red or green) or nonlinear (blue) fit.

Shimojo et al. (2017) used the four Japanese radio fluxes in order to study the evolution of microwave spectrum from 1957 to 2016, that is, over five and half solar cycles. They fit the logarithms of the four radio fluxes as a linear function of radio frequency in each month, and studied the solar cycle variation of the respective monthly intercepts and slopes. While the intercepts showed only a weak (in-phase) variation with the solar cycle, the slopes were found to vary by more than a factor of two in an opposite phase with the solar cycle. This shows that the radio spectrum varies systematically over the solar cycle and, at high activity, the longer wavelengths are enhanced relatively more than shorter ones. They also showed that the monthly spectra during the minimum-activity months of the four solar minima covered in the study, were almost exactly the same. Shimojo et al. (2017) concluded that the minimum-time (quiet-Sun) solar atmosphere above the chromosphere has not varied for half a century. These observations by Shimojo et al. (2017) have interesting consequences on the interpretation of our results, which will be discussed in Section 8.7.

As discussed in Section 8.5, some earlier papers have discussed the question of the nonlinear relation between the chromospheric and photospheric parameters (Foukal 1998; Preminger & Walton 2007; Tapping & Morgan 2017; Yeo et al. 2020; Clette 2021). It has also been suggested (see, e.g., Chatzistergos et al. 2022, and references there) that some of these relationships depend on the overall solar activity. Whether or not this finding is a related to our results, is still not clear. We have shown here that there is a clear long-term change between several photospheric, chromospheric and lower coronal parameters at yearly time resolution where hardly any nonlinearity is seen in their mutual relationships. In addition, we also see changes over the solar cycle where, as shown, for example, in the right panel of Fig. 3, the maxima in the F10.7/sunspot ratio are found typically around solar maxima. The nonlinear relation, for example, between plages and sunspots has been suggested to be due to the longer lifetime of plages (see, e.g., Preminger & Walton 2005, 2007). Models taking this into account (see, e.g., Yeo et al. 2020, and references therein) indeed find nonlinear relations and im-proved correlations between daily values of sunspots and plages but do not find long-term changes in their mutual relation. This suggests that the current observations are not directly related to those processes that produce the nonlinear relation between the photospheric and chromospheric parameters. In the next Section 8.7 we will present a novel physical interpretation for the observed long-term change in the mutual relationship between the parameters that originate from a wide range of altitudes in the solar atmosphere. This interpretation also naturally explains why the parameters at higher altitude systematically increase with respect to the lower-altitude parameters.

### 8.7. Interpretation: Shrinking due to weakening activity

We have shown that, as the overall solar activity weakens during the decay of the Modern Maximum, the solar parameters describing magnetic field elements of different field strength or being produced at a wide range of heights of solar atmosphere from the photosphere to the chromosphere and the low corona, vary differently. This leads to the observed long-term changes in the mutual relation of several of these parameters.

Both the strongest magnetic elements (sunspots) and the moderately strong elements of active regions are greatly reduced during the decay of the MM from cycle 19 to cycle 24. A small fraction of solar radio fluxes is generated (via gyroresonance) in sunspots, and is thus expected to be reduced by roughly the same rate as the sunspot number. However, a much larger fraction of radio fluxes is produced in active regions (plages) in the chromosphere and low corona via bremsstrahlung, which depends on local electron density. Since the radio fluxes decay less rapidly than sunspots, this suggests that the long-term decrease in sunspots (strong magnetic fields) proceeds somewhat faster than the decrease in density of active regions at chromospheric and low coronal altitudes.

With weakening solar activity, the number of moderately intense magnetic elements forming active regions is also decreasing. However, we found that the number of active regions is increasing in the long term with respect to the number of sunspots. While it is possible that this indicates that the distribution of magnetic elements of different field intensity is being changed during the overall weakening, it is also possible that the decreasing number of magnetic elements with sufficient intensity to be classified as an active-region element will lead to the fragmentation of the spatial distribution of these elements and to the separation of formerly contiguous active regions to a larger number of smaller active regions.

As noted above, the frequency of radio waves produced via bremsstrahlung depends on the plasma density of active regions. Accordingly, shorter (longer) radio waves are produced in more dense (rarefied, respectively) regions at a somewhat lower (higher) altitude in the solar atmosphere. We made here an interesting observation that the flux of the longer radio waves increased relative to the shorter waves during the decay of the MM. This change cannot be understood if only the number of active regions and related field lines is being reduced. However, it can be explained if the distribution of plasma along the active region field lines is changing at the same time as the total plasma density trapped in the solar atmosphere is decreasing due to the weakening solar activity.

Due to decreasing plasma density, the sources of all solar radio waves get to lower altitudes in order to adjust to the correct plasma density corresponding to their respective wavelength. When moving downward, the source of the longer radio waves at a higher altitude would have to move a relatively smaller step downward than the source of the shorter waves at a slightly lower altitude because of a larger volume compression (density increase) at the higher altitude in a canopy-structured system of closed magnetic field lines. Moreover, since moving downward in solar atmosphere (above the photosphere) implies moving to a cooler temperature, this also implies that the longer waves would cool relatively less than the shorter waves, which would enhance the emission of longer waves relative to shorter waves. Since the magnetic field in the chromosphere and low corona indeed has a canopy-structure and since the electrons follow the magnetic field lines, our results suggest that not only is the total plasma density decreasing but also the plasma height profile is slightly changing at these altitudes due to the weakening solar activity during the decay of the MM.

The strongest magnetic fields form the photospheric roots of the longest field lines, extending to high altitudes in the solar atmosphere (corona) and to large horizontal distances, some of them even connecting the two solar hemispheres. These field lines also determine the longest distance of closed field lines, that is, practically the source surface distance where the corona opens to the radially propagating solar wind. The reduction of strong magnetic elements would therefore lead to an effective shrinking of the solar coronal magnetic structure. It would also reduce the total amount of plasma in solar atmosphere and change the altitude distribution of plasma density in the solar atmosphere. We note that an earlier study by Virtanen et al. (2020) showed that the mismatch between the long-term evolution of the observed photospheric magnetic field (and the ensuing modeled coronal field) and the observed heliospheric magnetic field can be explained to high precision if the coronal source surface has indeed reduced due to the overall weakening during the last few decades.

As noted above, Shimojo et al. (2017) found that the minimum-time radio spectra remained almost the same during the four solar minima (after cycles 19-23) covered in their study, suggesting that the quiet-Sun atmosphere above the chromosphere has not varied during this time. This does not contradict with our observations. In fact, as we have noted in this paper (see, e.g., Fig. 3 and related discussion in Section 4), the observed long-term change occurs via a series of humps around solar maximum times. So, the long-term change observed here is intimately connected with a change in the active Sun rather than in the quiet Sun, which indeed remains fairly constant. Finally, we note that the largest relative changes between the different radio fluxes are seen to start in the 1990s (see Fig. 5), in a fair temporal proximity to those changes noted in Virtanen et al. (2020).

## 9. Conclusions

Five independent series of radio flux observations at different frequencies show that there was a systematic, long-term relative increase in radio fluxes with respect to the sunspot number from the 1970s until 2010s. This confirms the physical validity of these changes and excludes the recent claim (Clette 2021) that there was a step-like inhomogeneity in the 10.7 cm radio flux. The suggested 1980 jump (Clette 2021) was not a step-like increase, but only the first in a series of several short-term (1-2-year) humps, most of which occur during solar maxima. These humps contribute to the long-term increase in the radio flux - sunspot ratio observed in this study.

We used yearly averages when studying the mutual relations of the different solar parameters. When using yearly averages, these relations are only weakly nonlinear, while the daily av-erages between, for example, photospheric and chromospheric parameters typically depict considerably larger inhomogeneities (Foukal 1998). We showed that a nonlinear model yields essentially the same results as a linear model for the studied relations, which excludes the possibility that nonlinearity has a significant influence in these relations. This verifies the observed long-term changes in the various relations when using a linear model with yearly averages.

We found that also the number of active regions increased with respect to the sunspot number during the same time interval. We suggested that this increase could be due to the fragmentation of large active regions to a greater number of smaller regions due to the overall weakening of solar activity. On the other hand, the number of active regions did not increase with respect to 10.7 cm radio flux, which is expected since a large majority of 10.7 cm radio flux is produced in non-sunspot active regions. Still, this result is quite reassuring for the credibility of the obtained results.

The longer-wave radio fluxes were found to increase relative to the shorter waves from the 1960s to 2010s. We argued that this indicates an effective change in the altitude distribution of plasma density in the canopy-type magnetic topology of the chromosphere and low corona where the volume reduction (density increase) of descending plasma is larger at a higher altitude, closer to the canopy than to its body. This leads to a relatively shorter downfall and a smaller temperature decrease for the source region of longer radio wavelengths, leading to a relative increase in emission rate, as compared to shorter wavelengths. These changes are very likely related to the earlier-found shrinking of the solar corona with the weakening of solar activity during the decay of the MM, which reduces the longest closed field lines (and the related source surface of solar wind), reducing the overall plasma density in the solar atmosphere and changing its altitude distribution.

Most of the observed changes took place from the 1970s until 2010s, when the most recent Gleissberg cycle of solar activity, the Modern Maximum, was subsiding and the overall solar activity was weakening. We note that changes opposite to those reported here have very likely started growing already, because the activity in cycle 25 recently surpassed cycle 24 (for current situation, see SILSO web page[7]). This may be the first step toward the grand maximum of the 21st century, the Future Maximum (FM), whose cycle heights and hemispheric asymmetry have been predicted by a recent model, which explains the Gleissberg cyclicity in terms of an oscillating relic field in the Sun (Mursula 2023). It is also very likely that long-term changes opposite to those reported here already occurred in the growth phase of the MM from the 1920s to 1950s (and other, earlier Gleissberg cycles), and that these changes are a regular part of Gleissberg cyclicity, and thus occur during all Grand maxima.

*Acknowledgements.* [8]). The Research Council of Finland is acknowledged for support to T. Asikainen (projects no. 321440 and no. 357249) and to I. Tähtinen (project no. 321440). A. R. Yeates was supported by UK STFC grant ST/W00108X/1. Authors acknowledge the international team no. 475 "Modeling Space Weather And Total Solar Irradiance Over The Past Century" supported by the International Space Science Institute (ISSI), Bern, Switzerland and ISSI-Beijing, China. We acknowledge the LISIRD server (https://lasp.colorado.edu/lisird/) for providing links to the NOAA F10.7 cm index, to the Japanese (Toyokawa and Nobeyama) radio fluxes and to the Bremen MgII composite index. The Nobeyama Radio Polarimeters (NoRP) are operated by Solar Science Observatory, a branch of National Astronomical Observatory of Japan, and their observing data are verified scientifically by the consortium for NoRP scientific operations. Dr. M. Weber is acknowledged for constructing and maintaining the web site for the Bremen MgII composite index. The recent Penticton 10.7 cm data were retrieved from the NRCan server (https://www.spaceweather.gc.ca/forecast-prevision/solar-solaire/solarflux/sx-5-flux-en.php) which is served by the Solar Radio Monitoring Program operated jointly by the National Research Council Canada and Natural Resources Canada with support from the Canadian Space Agency. We acknowledge the service of daily, monthly and yearly total sunspot data (version 2) at the World Data Center SILSO, Royal Observatory of Belgium, Brussels (https://www.sidc.be/SILSO/). Active region data determined from National Solar Observatory (NSO) synoptic Carrington maps were downloaded from the Solar Dynamo Harvard Dataverse (https://doi.org/10.7910/DVN/Y5CXM8). The NSO is operated by the Association of Universities for Research in Astronomy (AURA), Inc. under a cooperative agreement with the National Science Foundation. We also acknowledge Dr. F. Clette for additional information on Fig. 24 in Clette (2021).

[7] https://www.sidc.be/SILSO/
[8] https://www.sidc.be/SILSO/